\documentclass[12pt,twoside]{article}
\usepackage{graphicx}
\usepackage{dcolumn}
\usepackage{bm}
\usepackage{psfrag}
\usepackage{amsmath}
\usepackage{latexsym}

\newcommand{\be}{\begin{equation}}
\newcommand{\ee}{\end{equation}}
\newcommand{\bee}{\begin{eqnarray}}
\newcommand{\eee}{\end{eqnarray}}

\date{}

\begin{document}

\centerline{}

\centerline{}

\centerline {\Large{\bf Shape vibrations of topological fermions}}

\centerline{}

\centerline{\bf {Manfried Faber}}

\centerline{}

\centerline{Technische Universit\"at Wien, Wiedner Hauptstr. 8-10,}

\centerline{A--1040 Vienna, Austria}

\centerline{}

\centerline{\bf {Alexander Kobushkin}}

\centerline{}

\centerline{Bogolyubov Institute for Theoretical Physics, Metrologicheskaya str. 14-B,}

\centerline{03680 Kiev, Ukraine}

\centerline{Physical and Technical National University of Ukraine KPI,}

\centerline{Prospect Pobedy 37, 03056 Kiev, Ukraine}

\centerline{}

\centerline{\bf {Mario Pitschmann}}

\centerline{}

\centerline{Technische Universit\"at Wien, Wiedner Hauptstr. 8-10,}

\centerline{A--1040 Vienna, Austria}

\centerline{}

\begin{abstract}
We analyze the model of topological fermions, where charged fermions are treated as topological solitons. We discuss vibrations of soliton shapes. It is shown that depending on the power of the potential term (discrete parameter $m$) of the model Lagrangian the spectrum of normal mode frequencies can be discrete (for $m=1$) or continuous (for $m\ge 2$).
\end{abstract}

{\bf PACS:} 05.45.Yv\\

{\bf Keywords:} solitons, vibrations, hedgehog

\section{Introduction}
The well-known success of the Skyrme model to the description of short-range forces and properties of strongly coupled particles makes it worthwhile to extend the ``skyrmion philosophy'' to long-range forces and physics of electrically coupled particles. The so-called model of topological fermions (MTF) \cite{Manfried,FaberKob} proposes a realization of such an idea.

The model has three independent degrees of freedom parameterizing an SU(2) field
\be
\label{I.2}
Q(x)=\cos\alpha(x) + i\vec \sigma\vec n(x)\sin\alpha(x),
\ee
where $\vec \sigma$ are the Pauli matrices and $\vec n(x)$ is a three-dimensional unit vector in internal (``colour'') space~\footnote{We use the summation convention that any capital latin index that is repeated in a product is automatically summed from 1 to 3. The arrows on variables in the internal ``colour'' space indicate the set of 3 elements $\vec{q}=(q_1, q_2, q_3)$ or $\vec{\sigma}=(\sigma_1, \sigma_2, \sigma_3)$ and $\vec{q} \vec{\sigma}= q_K \sigma_K$. We use the wedge symbol $\wedge$ for the external product between colour vectors $(\vec{q}\wedge\vec{\sigma})_A=\epsilon_{ABC}q_B \sigma_C$. For the components of vectors in physical space ${\mathbf x}=(x,y,z)$ we employ small latin indices, $i,j,k$ and a summation convention over doubled indices, e.g. $({\mathbf E} \times {\mathbf B})_i = \epsilon_{ijk} E_j B_k$. Further we use the metric $\eta = \mathrm{diag}(1,-1,-1,-1)$ in Minkowski space.}. Due to the constraint $|\vec n(x)|=1$ this vector has two independent degrees of freedom. Both fields, $\alpha(x)$ and $\vec n(x)$, are functions of the Minkowski coordinates  $x^\mu=(ct,x,y,z)$.

The Lagrangian density of the MTF reads
\be\label{MTFlag}
\mathcal L=-\frac{\alpha_f\hbar c}{4\pi}\left(\frac14\vec R_{\mu\nu}\cdot\vec R^{\mu\nu}+\Lambda(q_0)\right),
\ee
where $\vec R^{\mu\nu}$ is the curvature tensor
\be\label{connection}
\vec R^{\mu\nu}=\vec\Gamma^\mu\wedge\vec\Gamma^\nu, \qquad
\text{with the connection} \qquad
\vec \Gamma^\mu=\frac1{2i}\mathrm{Tr}(\vec\sigma \partial^\mu QQ^\dag).
\ee
The potential term is given by
\be\label{model_potential}
\Lambda(q_0)=\frac1{r_0^4}\left(\frac{\mathrm{Tr}Q}{2}\right)^{2m}=
\frac1{r_0^4}\cos^{2m}\alpha(x), \quad m=1,2,3,\dots
\ee
The model contains two parameters, the fine-structure constant, $\alpha_f
$, and a dimensional parameter $r_0$.

Note that the ``curvature term'' $-\frac14\vec R_{\mu\nu}\cdot\vec R^{\mu\nu}$ is proportional to the Skyrme term, but the so-called kinetic term of the Skyrme model does not enter the Lagrangian density (\ref{MTFlag}) in order to allow for electromagnetic fields and forces \cite{FaberKob}.

Due to its Lagrangian density the MTF has different properties than the Skyrme model at $r\to\infty$ \cite{Manfried,FaberKob}. In the Skyrme model the chiral field $U$ approaches the trivial configuration, $U\to 1$. In the MTF the field configuration for $r \to \infty$ is determined by the minima of the potential 
characterized by  $\alpha(x)=\frac{\pi}2$ and arbitrary direction of $\vec n$,
\be\label{Q_infty}
Q(x)=i\vec \sigma \vec n(x) \qquad \text{at}\qquad r\to\infty.
\ee
As a result the $Q$ field can form a hedgehog configuration and the field $\alpha(x)$ describes the profile of a charged soliton with properties of an electron, whereas the field $\vec n(x)$ is related to the dual electromagnetic field strength \cite{Manfried,FaberKob} by
\be\label{EMfield}
^{\ast} f_{\mu\nu}(x)=-\frac{e_0}{4\pi\varepsilon_0 c}[\partial_\mu \vec n(x)\wedge\partial_\nu \vec n(x)]\cdot \vec n(x).
\ee
The field strength $f_{\mu\nu}$ reads $f_{\mu\nu}=-\frac12\epsilon_{\mu\nu\rho\sigma}{^{\ast} f^{\rho\sigma}}$ with $\epsilon^{0123}=1$.

The model has two types of excitations, which are related to $\vec n$ and $\alpha$ degrees of freedom and have different physical meaning.

One type (connected with fluctuations of the field $\vec n(x)$) is realized as electromagnetic field evolved by the charge in the wave zone (where $\alpha\to \frac{\pi}2$). Some properties of such fluctuations were already studied, appropriate  classical equations of motion for the field $\vec n$ were derived \cite{FaberKob} and explicit solutions of these equations of motion, which behave like electromagnetic waves, were found \cite{BorFabKob}.

The subject of the present paper is to study another type of MTF excitations which is generated by fluctuations of the field $\alpha(x)$ and realized as vibrations of the soliton shape.

The paper is organized as follows. In Section~\ref{sec:2} we derive the Lagrangian density for small $\alpha$-fluctuations. Than, in Section~\ref{sec:3}, we calculate normal modes of the fluctuations. We find that the spectrum of the mode frequencies is very different for $m=1$ and $m\neq 2$, where $m$ is the power in the potential term (\ref{potential}). The conclusions are given in Section~\ref{sec:conclusions}.
\section{$\alpha$-fluctuation in the MTF Lagrangian density \label{sec:2}}
We will start from the second order variation terms of the MTF Lagrangian density
\begin{equation}
\begin{aligned}
\delta {\cal L}
=&-\tfrac{\alpha_f\hbar c}{4\pi}\; \delta \left(\tfrac{1}{4} \vec{R}_{\mu\nu}\vec{R}^{\mu\nu} + \Lambda\right) \\
=&\;\tfrac{\alpha_f\hbar c}{4\pi}\left[
\tfrac12\vec{\zeta}^{\;2} q_0 \; \partial_{q_0} \Lambda
-\tfrac12(\vec{\zeta}\vec{q}\,)^2 \partial_{q_0}^2 \Lambda
-(\partial_\mu\vec{\zeta} \wedge \partial_\nu\vec{\zeta}\,)\vec{R}^{\mu\nu}+(\partial_\mu\vec{\zeta} \wedge \vec{\zeta})\;
(\vec{\Gamma}_\nu\wedge\vec R^{\mu\nu})\right.\\
&\left.+\tfrac{1}{2}(\partial_\mu\vec{\zeta} \wedge \vec{\Gamma}^\mu)
(\partial_\nu\vec{\zeta} \wedge \vec{\Gamma}^\nu)
-\tfrac{1}{2}(\partial_\mu\vec{\zeta} \wedge \vec{\Gamma}_\nu)
(\partial^\mu\vec{\zeta} \wedge \vec{\Gamma}^\nu)\right],
\end{aligned}
\label{carte}
\end{equation}
where $\vec{\zeta}=\vec{\zeta}(t,\mathbf x)$ are three independent variation parameters introduced by
\begin{equation}\label{varyQ}
\begin{aligned}
Q \rightarrow Q^\prime &= e^{i\vec{\sigma}\vec{\zeta}} Q =\\
&=\left[\left(1-\tfrac12\zeta^2\right) + i \vec{\sigma}\vec{\zeta}\right](q_0+i\vec{q}\vec{\sigma})=\\
&= \left[ q_0(1-\tfrac12\zeta^2) - \vec{\zeta}\vec{q} \right] + i \vec{\sigma} \left[(1-\tfrac12\zeta^2)\vec{q} + q_0 \vec{\zeta} + \vec{q}\wedge\vec{\zeta} \right].
\end{aligned}
\end{equation}
In Eq.~(\ref{carte}) terms linear in $\vec{\zeta}$ vanish due to the equation of motion.

To simplify calculations we introduce in (\ref{carte}), (\ref{varyQ}) and later on the notations $q_0(x)=\cos \alpha(x)$ and $\vec q(x)=\vec n(x)\sin \alpha(x)$ and use spherical coordinates $\theta$ and $\phi$ in colour space.

We will also use the spherical coordinates, $r,\vartheta,\varphi$ for the physical space and specify the hedgehog soliton by $\vec n=\frac{\vec r}{r}$. One gets
\be
\label{unit}
\vec e_r=
\vec n,
\qquad\partial_\vartheta\vec n=\vec e_\vartheta,
\qquad\frac{\partial_\varphi\vec n}{\sin\vartheta}=\vec e_\varphi.
\ee
By rotation of $\vec{e}_\theta$ and $\vec{e}_\xi$ and $\vec{e}_\phi$ with angle $\alpha$ we get
\be\label{rotation}
\vec{e}_\xi= \cos \alpha \, \vec e_\theta - \sin \alpha \, \vec e_\phi,
\qquad \vec e_\eta=\sin \alpha \, \vec e_\theta + \cos \alpha \,  \vec e_\phi
\ee
with
\be\label{1}
(\vec e_\xi\wedge\vec e_\eta)=\vec n\qquad\text{and}\qquad
(\partial_\vartheta\vec n\wedge\partial_\varphi\vec n)\cdot\vec n=\sin\vartheta.
\ee
In spherical coordinates the covariant and contravariant components of the connection and the curvature tensor (\ref{connection}) read
\begin{equation}\label{sphericalVecPots}
\begin{aligned}
\begin{array}{ll}
\vec{\Gamma}_r = \alpha^\prime(r) \, \vec n ,&\vec\Gamma^r =-\alpha^\prime(r) \vec n,\\
\vec{\Gamma}_\vartheta =
\sin \alpha \, \vec{e}_\xi , &\vec\Gamma^\vartheta = -\frac{1}{r^2} \sin\alpha \vec e_\xi,\\
\vec{\Gamma}_\varphi =
\sin\vartheta \sin \alpha \, \vec{e}_\eta ,&\vec\Gamma^\varphi = -\frac{1}{r^2 \sin \vartheta} \sin\alpha \vec e_\eta\\
\vec R_{\vartheta \varphi} = \sin\vartheta \sin^2 \alpha \, \vec n,&
\vec R^{\vartheta \varphi} = \frac{\sin^2 \alpha}{r^4 \sin\vartheta} \, \vec n,\\
\vec R_{\varphi r} = \sin\vartheta \sin \alpha \, \alpha^\prime \, \vec{e}_\xi,\qquad \mbox{}&
\vec R^{\varphi r} = \frac{\sin \alpha \, \alpha^\prime}{r^2 \sin\vartheta} \, \vec{e}_\xi,\\
\vec R_{r \vartheta} = \sin \alpha \, \alpha^\prime \, \vec{e}_\eta,&
\vec R^{r \vartheta} = \frac{\sin \alpha \, \alpha^\prime}{r^2} \, \vec{e}_\eta.
\end{array}
\end{aligned}
\end{equation}

After some algebra one arrives at
\begin{equation}\label{secondorder}
\begin{aligned}
\delta {\cal L}
=&\;\tfrac{\alpha_f\hbar c}{4\pi}\left[\tfrac12\vec{\zeta}^{\,2} \cos \alpha \; \partial_{q_0} \Lambda
-\tfrac12\zeta_r^2 \sin^2 \alpha \; \partial_{q_0}^2 \Lambda-\tfrac{2\sin^2 \alpha}{r^4 \sin\vartheta}(\partial_\vartheta \vec{\zeta} \wedge \partial_\varphi \vec{\zeta}\,) \cdot \vec n\right.\\
&\left.
-\tfrac{2\sin \alpha \, \alpha^\prime}{r^2 \sin\vartheta}(\partial_\varphi \vec{\zeta} \wedge \partial_r \vec{\zeta}\,) \cdot \vec{e}_\xi
-\tfrac{2\sin \alpha \, \alpha^\prime}{r^2}(\partial_r \vec{\zeta} \wedge \partial_\vartheta \vec{\zeta}\,) \cdot \vec{e}_\eta+\tfrac{2\sin^2 \alpha \; \alpha^\prime}{r^2}
(\partial_r\vec{\zeta} \wedge \vec{\zeta}) \; \vec n\right.\\
&\left.
+\tfrac{\sin \alpha}{r^2}(\alpha^{\prime 2} + \tfrac{\sin^2 \alpha}{r^2})
[(\partial_\vartheta\vec{\zeta} \wedge \vec{\zeta}) \; \vec{e}_\xi
+\tfrac{1}{\sin\vartheta}(\partial_\varphi\vec{\zeta} \wedge \vec{\zeta}\,)\; \vec{e}_\eta]\right.\\
&\left.+\tfrac{1}{2}\{\alpha^\prime (\partial_r\vec{\zeta} \wedge \vec n)
+\tfrac{\sin \alpha}{r^2} [
(\partial_\vartheta\vec{\zeta} \wedge \vec{e}_\xi)
+\tfrac{1}{\sin\vartheta}(\partial_\varphi\vec{\zeta} \wedge \vec{e}_\eta)] \}^2\right.\\
&\left.+\tfrac{\sin^2 \alpha}{r^2}(\partial_\mu\vec{\zeta} \cdot \vec n)(\partial^\mu\vec{\zeta} \cdot \vec n)
+\tfrac{1}{2} (\alpha^{\prime 2} + \tfrac{\sin^2 \alpha}{r^2})
[(\partial_\mu\vec{\zeta} \; \vec{e}_\xi)
(\partial^\mu\vec{\zeta} \; \vec{e}_\xi)
+ (\partial_\mu\vec{\zeta} \; \vec{e}_\eta)
(\partial^\mu\vec{\zeta} \; \vec{e}_\eta)]\right].
\end{aligned}
\end{equation}
Now let us consider ``$\alpha$-fluctuations'', which correspond to the following choice of the parameter $\vec{\zeta}$
\be\label{alpha}
\vec{\zeta}(t,\mathbf x)=\phi(t,r)\vec n, \qquad \vec{n}=\frac{\mathbf x}{|\mathbf x|}\,.
\ee
Then the fluctuation of the Lagrangian density reads
\be\label{finally}
\begin{aligned}
\delta {\cal L}
=&\;\frac{\alpha_f\hbar c}{4\pi}\left[\frac{\sin^2 \alpha}{r^2}\partial_\mu\phi\partial^\mu\phi
-\phi^2\left\{\frac{\sin\alpha\sin3\alpha}{r^4} +\frac{\cos2\alpha}{r^2}\,\partial_\mu\alpha\partial^\mu\alpha-
\frac{2\sin2\alpha}{r^3}\,\partial_\mu r\partial^\mu\alpha+
\right.\right.\\
&\left.\left.+
\frac{\sin2\alpha}{r^2}\,\Box\alpha+\frac{m}{r_0^4}\cos^{2m}\!\alpha\Big((2m-1)\tan^2\!\alpha-1
\Big)\right\}\right].
\end{aligned}
\ee
Using the equation of motion
\be\label{EoM}
\frac{m}{r_0^4}\cos^{2m}\!\alpha=-\frac{\sin^2\alpha\cos\alpha}{r^4}+2\frac{\sin\alpha}{r^3}\partial_\mu r\,\partial^\mu\alpha-\frac{\cos\alpha}{r^2}\,\partial_\mu\alpha\partial^\mu\alpha-\frac{\sin\alpha}{r^2}\,\Box\alpha
\ee
the expression (\ref{finally}) reduces to
\be\label{finally1}
\begin{aligned}
\delta {\cal L}
=&\;\frac{\alpha_f\hbar c}{4\pi}\left\{\frac{\sin^2 \alpha}{r^2}\,\partial_\mu\phi\partial^\mu\phi
-\phi^2\left[
\frac{\sin^2\alpha(1+3\cos2\alpha)}{2r^4} -
\frac{\sin2\alpha}{r^3}\,\partial_\mu r\partial^\mu\alpha\right.\right.\\
&\left.\left.-\frac{\sin^2\alpha}{r^2}\,\partial_\mu\alpha\partial^\mu\alpha+
\frac{\sin2\alpha}{2r^2}\,\Box\alpha+\frac{m}{r_0^4}\cos^{2m}\!\alpha(2m-1)\tan^2\!\alpha
\right]\right\}.
\end{aligned}
\ee
Introducing a new field variable
\be\label{variable}
\Phi=\frac{\sqrt2\sin\alpha}{r}\phi
\ee
one arrives at the final expression for the fluctuating Lagrangian density
\be\label{MarioLagra}
\delta {\cal L}
=\;\frac{\alpha_f\hbar c}{4\pi}\left\{\frac12\,\partial_\mu\Phi\partial^\mu\Phi
-\Phi^2\left[
\frac{1+3\cos2\alpha}{4r^2} +\frac{m(2m-1) r^2}{2r_0^4}\cos^{2m-2}\!\alpha
\right]\right\}.
\ee
\section{Normal modes of fluctuations \label{sec:3}}
\subsection{Fluctuation potential}
From (\ref{MarioLagra}) we get the Lagrangian for the $\alpha$-fluctuations. It has the standard form
\be\label{StandardForm}
\begin{aligned}
L[\Phi]&=T[\Phi]-U[\Phi],\qquad T[\Phi]=\textstyle{\frac12}\int d^3x\;\Dot\Phi^2,\\
U[\Phi]&=\int d^3x\left[\textstyle{\frac12}(\nabla\Phi)^2+\Phi^2V(r)\right]\\
&=\int d^3x\;\Phi\left[-\textstyle{\frac12}\Delta+V(r)\right]\Phi,\\
V(r)&=\displaystyle\frac{1+3\cos2\alpha}{4r^2} +\frac{m(2m-1) r^2}{2r_0^4}\cos^{2m-2}\!\alpha=\frac{v(\rho)}{r^2_0} ,
\end{aligned}
\ee
where $\rho=r/r_0$.

\begin{figure}[h]
\centering
\includegraphics[width=0.475\textwidth]{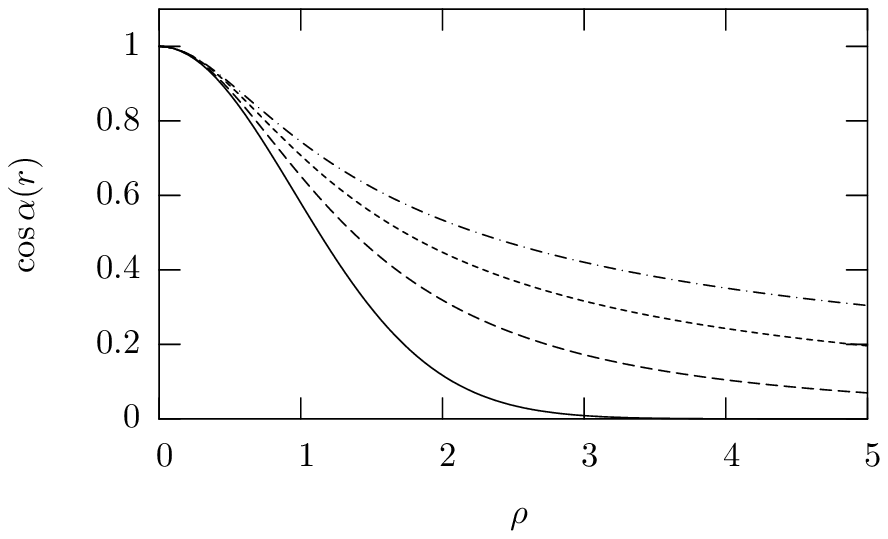}
\includegraphics[width=0.475\textwidth]{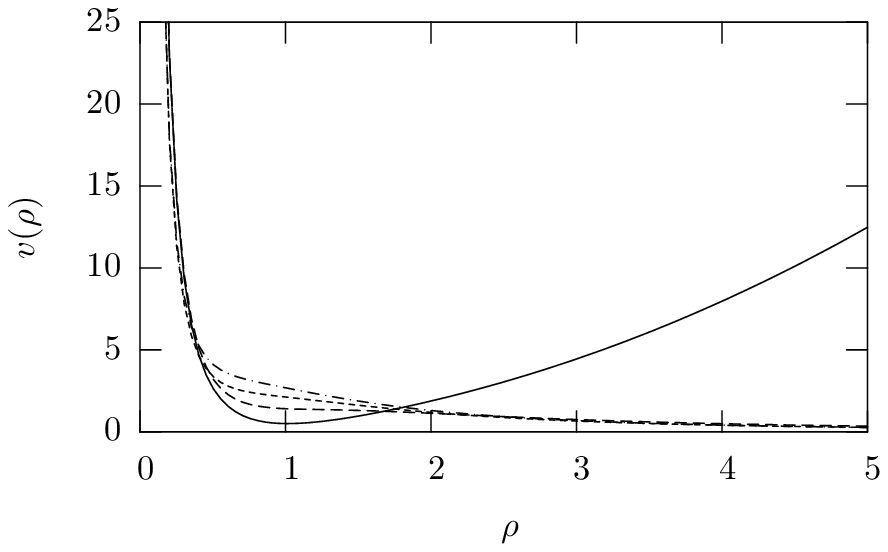}
\caption{At the left panel: $\cos\alpha(r)$ for $m=1$ (solid line), $m=2$ (long-dashed line), $m=3$ (short-dashed line) and $m=4$ (dot-dashed line). At the right panel: the potential $v$ in dependence on  $\rho=r/r_0$.}
\label{fig:profiles}
\end{figure}

To calculate the potential $V(r)$ one needs to know the profile function $\alpha(r)$. The solutions for the profile function with different $m$ are discussed in Appendix~\ref{sec:Appedix1}. The behaviour of  $\cos\alpha(r)$ for $m=$1,2,3 and 4 is displayed in Figure~\ref{fig:profiles}.

From Eq.~(\ref{StandardForm}) one immediately learns that for $m=1$  the fluctuation potential increases as $V(r)\sim r^2$ at $r\to \infty$, while for $m\ge 2$ it decreases as $V(r)\sim r^{-2}$, see Figure~\ref{fig:profiles}. Outside the region 0.3$\;<\rho<\;$2 the potential is very similar for all $m\ge 2$.

For $m=$2 and 3 we can give explicit expressions for the potential
\be\label{potential}
v(\rho)=\left\{
\begin{array}{ll}
\displaystyle\sqrt{\frac27}\cdot\frac{2-2\tilde\rho^2+10\tilde\rho^4}{\tilde\rho^2(1+\tilde\rho^2)^2}, & m=2\\[.5cm]
\displaystyle\frac{2+\rho^2+14\rho^4}{\rho^2(1+\rho^2)^2}, & m=3
\end{array}
\right.
\ee
where $\widetilde \rho$ is defined in (\ref{casem=2}).
\subsection{Normal modes}
To find the normal modes of the fluctuations we expand $\Phi(t,\mathbf r)$ in an orthonormal and complete set of functions $\eta_i(\mathbf r)$ (see, e.g, \cite{Radjaraman})
\be\label{time_dep}
\Phi(t,\mathbf r)=\sum_i c_i(t)\eta_i(\mathbf r).
\ee
These functions  $\eta_i(\mathbf r)$ are determined by the Schr\"odinger-type equation
\be\label{DifEq}
\left[-\textstyle{\frac12}\Delta+V(r)\right]\eta_i(\mathbf{r})=\Omega^2_i\eta_i(\mathbf{r}).
\ee

Because the potential is very different for $m=1$ and $m \ge 2$ let us consider these two cases separately.
\begin{itemize}
\item \underline{$m=1$}

There is an infinite number of bound states. Separating angular and radial coordinates
$\eta_i(\mathbf{r})=\frac1r R_{nl}(r) Y_{ll_3}(\mathbf r/r)$ in Eq.~(\ref{DifEq}) we get for the radial wave function
\be\label{readialEq}
-\frac12 R_{nl}''+\left[\frac{l(l+1)}{2r^2}+V(r)\right] R_{nl}=\Omega^2_{nl} R_{nl},
\ee
where
\be\label{Pot1}
V(r)=\frac{1+3\cos 2\alpha(r)}{4r^2}+\frac{r^2}{2 r_0^4}=\frac{v(\rho)}{r_0^2}
\ee
and $\alpha(r)$ is given by (\ref{trial_1}).

From (\ref{readialEq}) and (\ref{Pot1}) it follows that for $l\gg 1$  the spectrum reduces to the spectrum of the three dimensional harmonic oscillator.

At $r\to 0$ Eq.(\ref{readialEq}) becomes
\be\label{r_to_0}
-\frac12R_{nl}''+\left[\frac{l(l+1)+2}{2r^2}\right] R_{nl}=0
\ee
with the solution
\be\label{sol_r_to_0}
R_{nl} \sim r^\xi,\qquad \xi=\frac12 +\sqrt{\frac14 +l(l+1) +2}.
\ee
At $r\to\infty$ it reduces to the asymptotic oscillator equation
\be\label{r_to_infty}
-\frac12R_{nl}''+ \frac1{2}\kappa r^2R_{nl}=0,\qquad \kappa=\omega^2=\frac1{r_0^4}
\ee
with the solution
\be\label{sol_r_to_infty}
2R_{nl}\sim r^\sigma e^{-\frac12\omega r^2}.
\ee

\begin{figure}[h]
\centering
\includegraphics[width=0.45\textwidth]{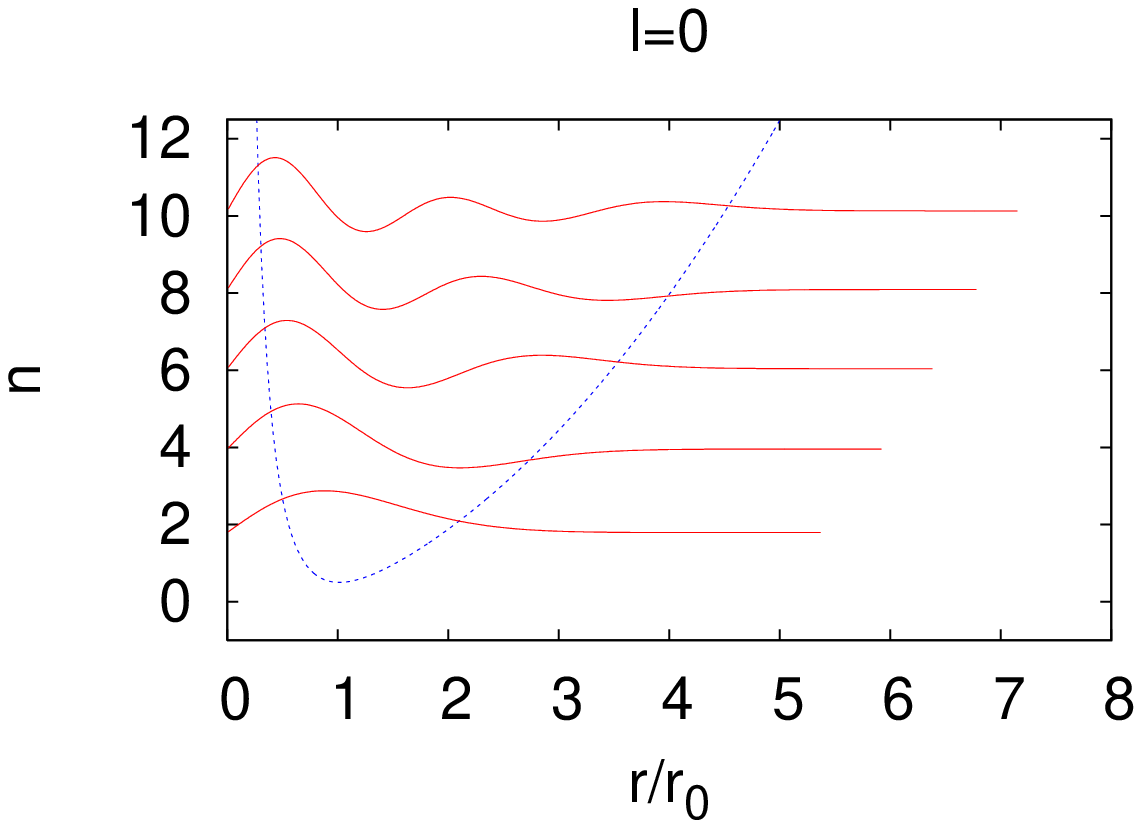}
\includegraphics[width=0.45\textwidth]{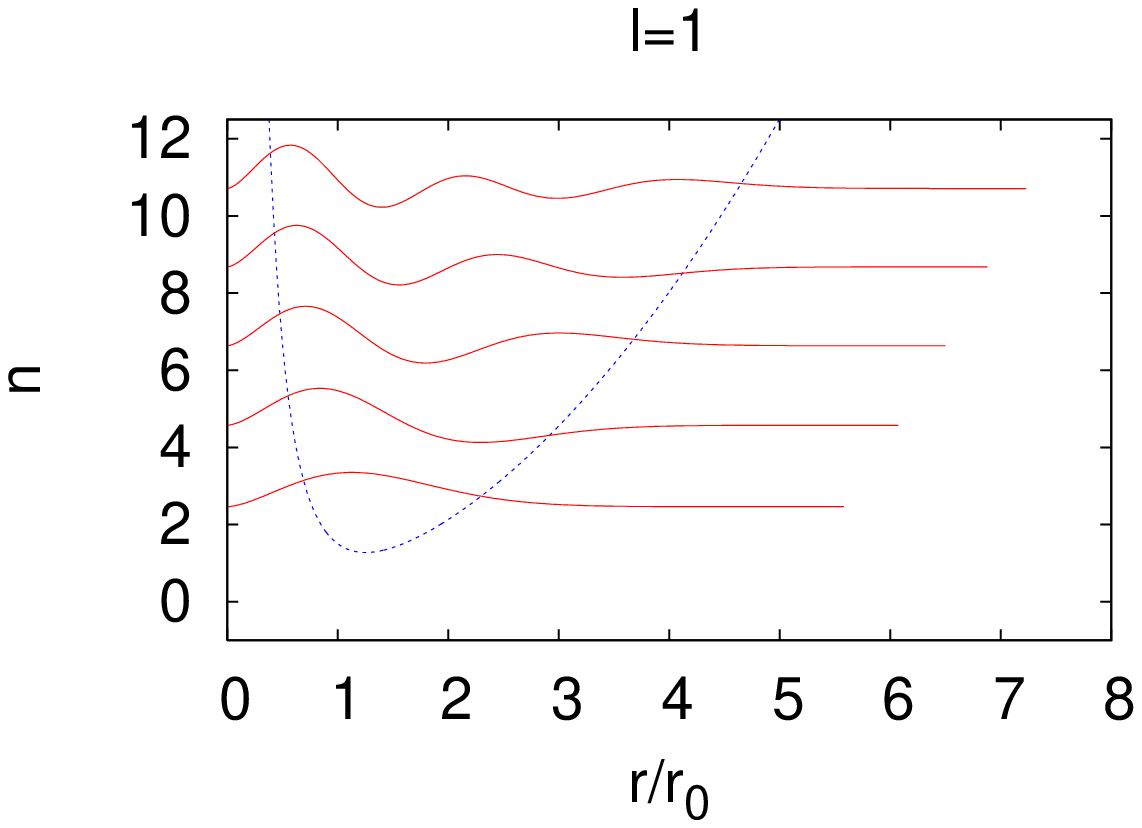}
\includegraphics[width=0.45\textwidth]{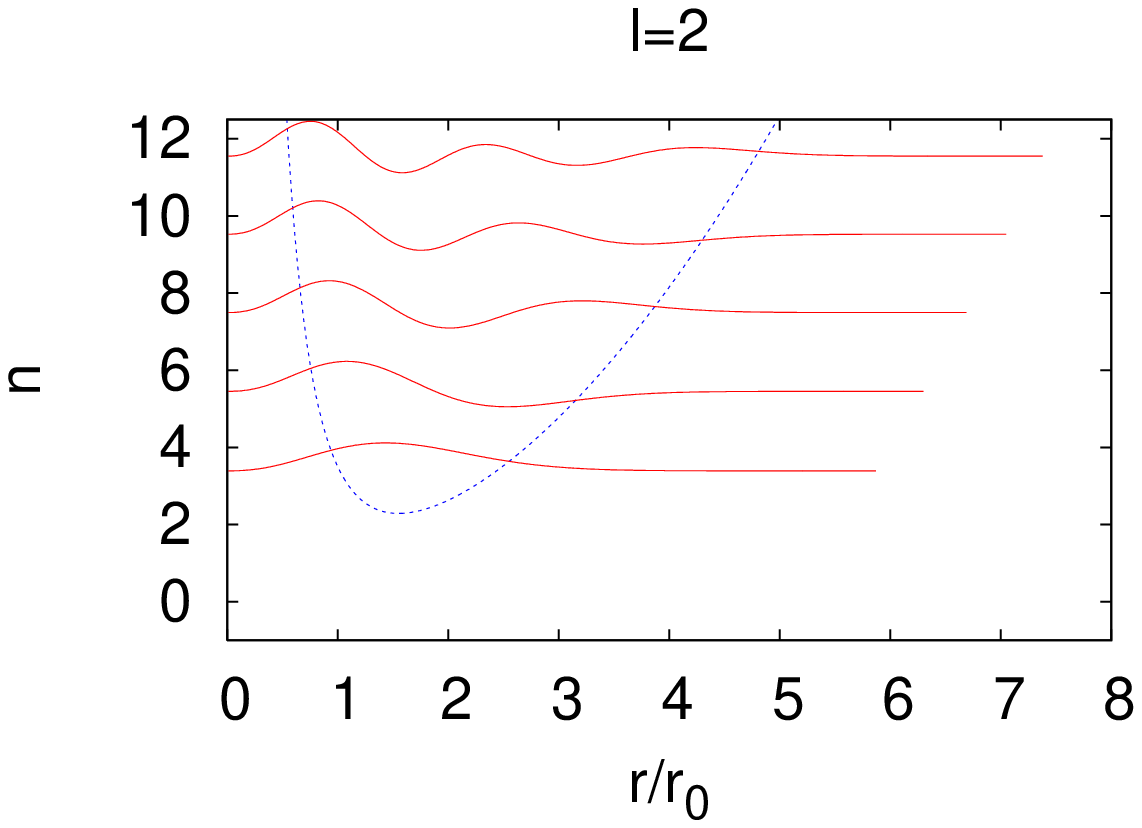}
\includegraphics[width=0.45\textwidth]{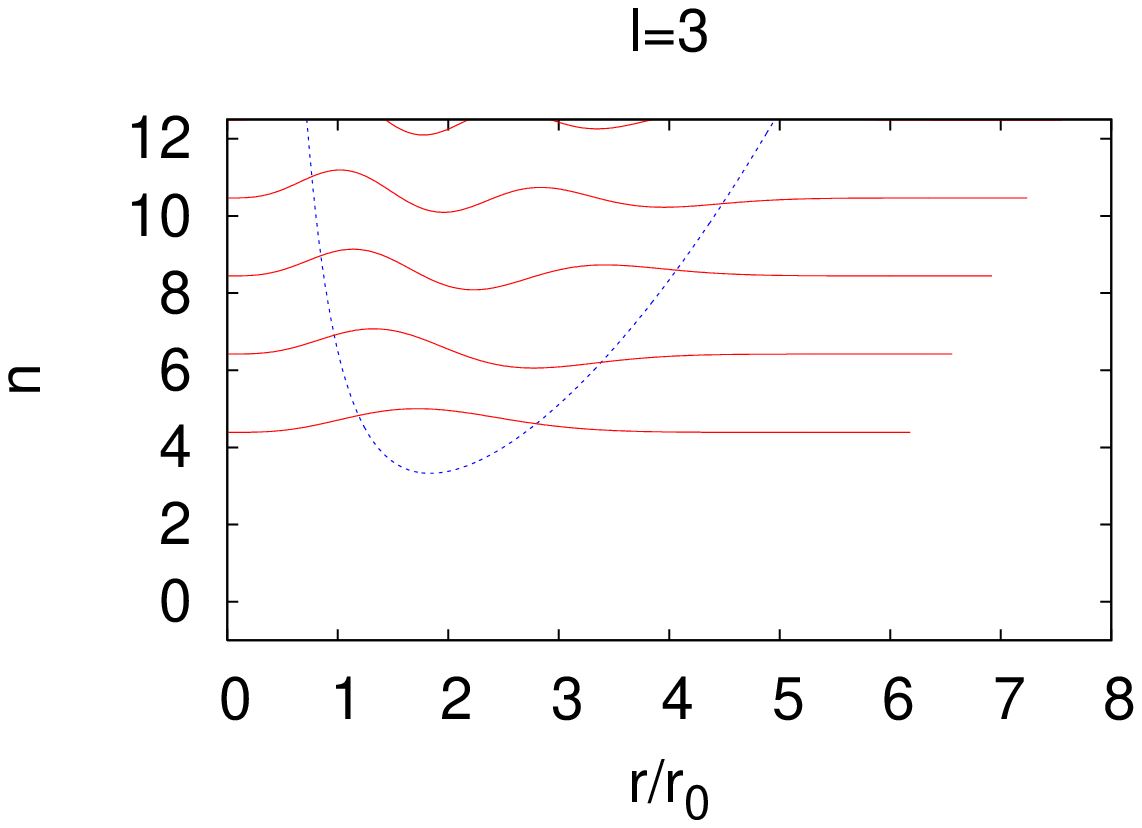}
\includegraphics[width=0.45\textwidth]{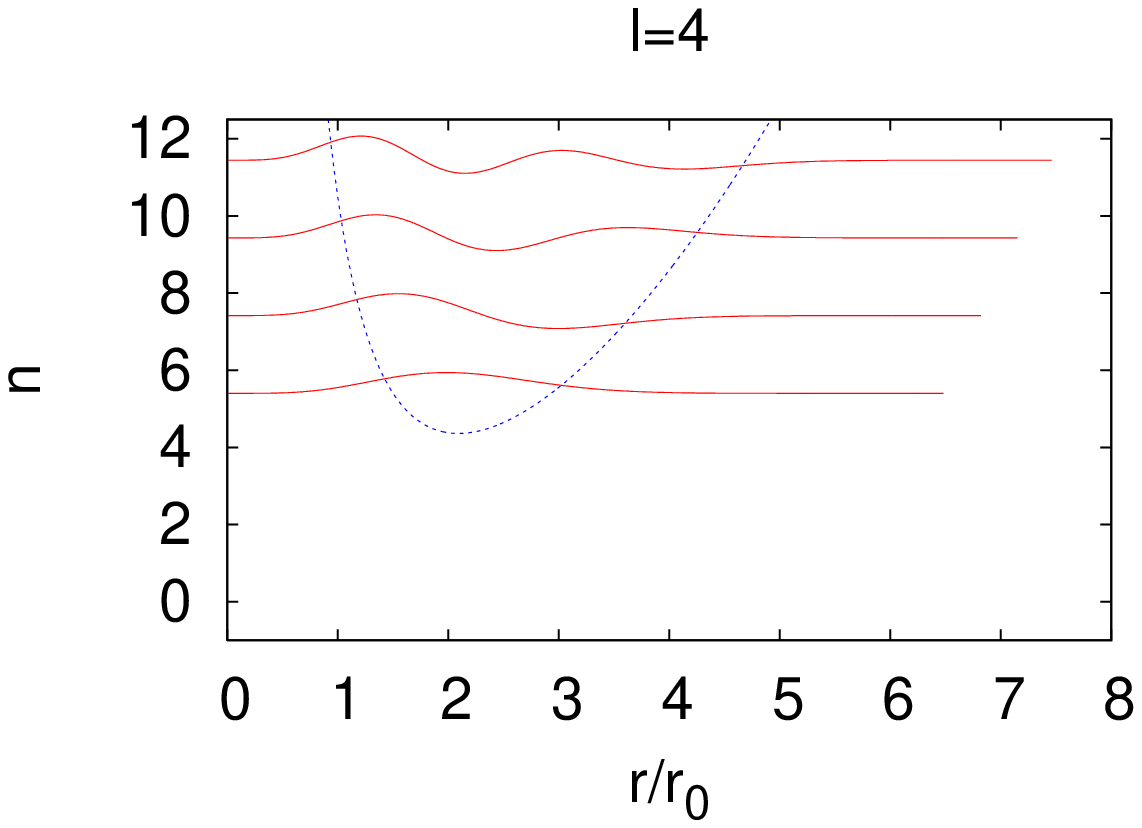}
\includegraphics[width=0.45\textwidth]{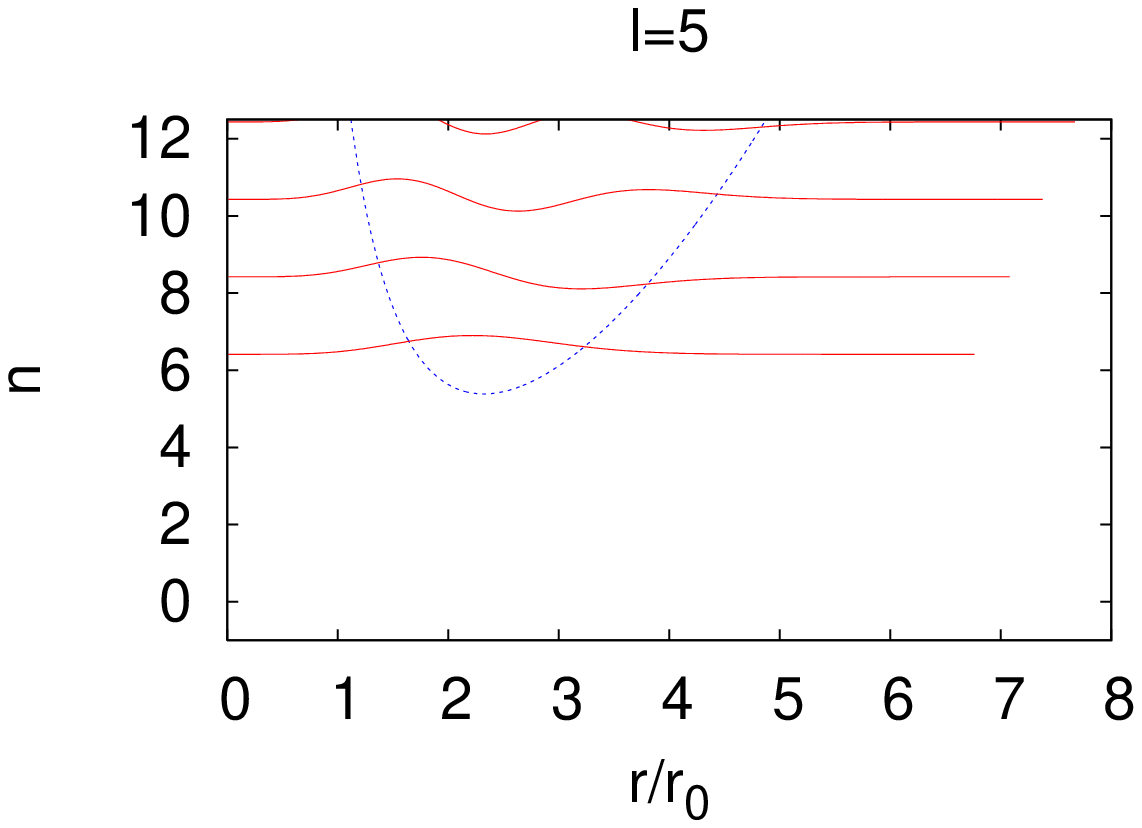}
\caption{The radial wave functions $R_{nl}$ for $m=1$. The radial wave functions are shifted by eigenvalues $\Omega^2_{nl}$. For comparison we show the sum of the potential and the centrifugal energy $\frac{l(l+1)}{2\rho^2}$.}
\label{fig:WFs}
\end{figure}

\begin{figure}[h]
\centering
\psfrag{l}{$l$}
\includegraphics[width=.8\textwidth]{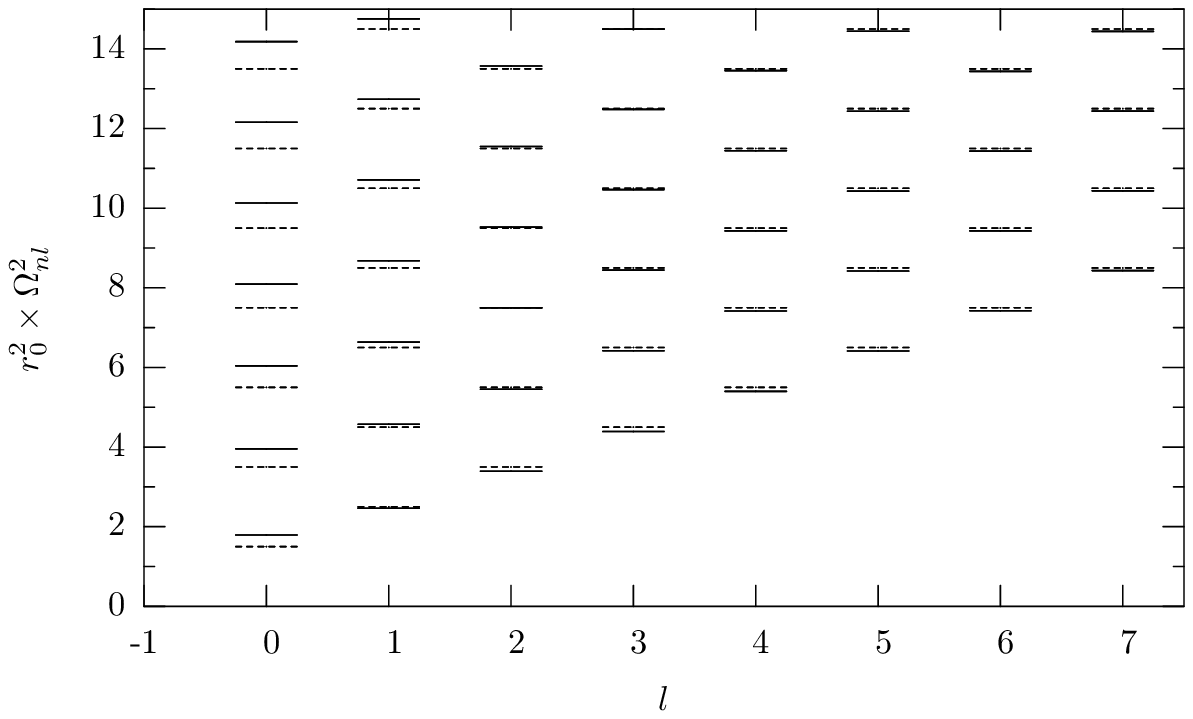}
\caption{The spectrum of Eq.~(\ref{DifEq}) for  $m=1$ (solid lines). For comparison we also show the spectrum $\widetilde\Omega_{nl}^2=\frac1{r_0^2}(2n+l-\frac12)$ of three-dimensional oscillator potential (dashed lines).}
\label{fig:eigenstates}
\end{figure}

The numerical solutions for the radial eigenfunctions are shown in Figure~\ref{fig:WFs} for different angular momentum quantum number $l$ and radial quantum number $n$. So the solutions of the Schr\"odinger equation (\ref{DifEq}) are characterized by three quantum numbers, $i=(n,l,l_3)$.  The eigenstates are obviously degenerate with respect to $l_3$.

In Figure~\ref{fig:eigenstates} the eigenvalues $\Omega_{nl}^2$ are compared with the spectrum $\widetilde\Omega_{nl}^2$ of the three dimensional oscillator.
\item\underline{$m\ge 2$}

For $m\ge 2$ the potential $V(r)$ is repulsive everywhere and there are no bound states. At $r\to \infty$ the solutions of Eq.~(\ref{DifEq}) are spherical waves
\be\label{spherical}
\eta_k=\frac{e^{\pm ikr}}{r}\;\qquad \text{and}\qquad \omega_k^2=\textstyle{\frac12}k^2.
\ee
The spectrum is continuous.
\end{itemize}
\section{Conclusions\label{sec:conclusions}}
We derive the Lagrangian for small fluctuations of the soliton profile around the hedgehog solution and discuss shape vibrations of the topological fermions. It is shown that at $r \ll r_0$ the potential term of the Lagrangian is repulsive and very similar for all values of the discrete parameter $m$ of the model. At $r \to \infty$ its behaviour differs for $m=1$ and $m\ge 2$. It grows as $r^2$ for $m=1$, while for $m\ge 2$ it decreases as $r^{-2}$. As a result the spectrum of normal frequencies is discrete for  $m=1$ and continuous for $m\ge 2$.\\[0.5cm]
{\bf Acknowledgements}\\[0.5cm]
This work was supported in part by ``Fonds zur F\"orderung der Wissenschaftlichen Forschung'' under contract P16910-N12.
\appendix
\section{Profile function\label{sec:Appedix1}}
The profile function is determined from the differential equation
\be\label{dif_eq}
\frac{d^2 q_0}{d\rho^2} + \frac{(1-q_0^2)q_0}{\rho^2}-
m\rho^2q_0^{2m-1}=0
\ee
supplemented by the following boundary conditions
\be\label{BCs}
\alpha(0)=0,\qquad \alpha(\infty)=\frac{\pi}2\,.
\ee
In (\ref{dif_eq}) one sets $q_0(r)=\cos\alpha(r)$ and $\Lambda(q_0)=q_0^{2m}(r)$.

At short distances, $\rho\ll 1$, Eq.~(\ref{dif_eq}) is fulfilled by
\be\label{small_distances}
q_0(r)\approx1-\kappa \rho^2
\ee
with arbitrary $\kappa$. The parameter  $\kappa$ is determined by the condition $q_0(\infty)=0$.

Now let us discuss the behaviour of $q_0$ at $r\to \infty$. This behaviour is very different for $m=1$ and $m\ge 2$. Thus we will consider these two cases separately. Note, that for $m=2$ and 3 there are exact analytical solutions, for other $m$ only approximate solutions do exist.\\[-0.5cm]

\underline{For $m=1$} one gets the asymptotic equation
\be\label{dif_eq_asymp_1}
\frac{d^2 q_0}{d\rho^2}-\rho^2q_0=0 \qquad \text{at}\qquad \rho\to \infty
\ee
with solution $q_0\sim \rho^{-\frac12} e^{-\frac12 \rho^2}$. To connect the solutions (\ref{small_distances}) and (\ref{dif_eq_asymp_1}) smoothly one can use the following trial function
\be\label{trial_1}
\cos\alpha(r)=\frac{ e^{-\frac12 \rho^2}}{\sqrt[4]{1+\kappa_0 \rho^2}}.
\ee
The variation parameter $\kappa_0=0.206796$ is determined from the minimum of the energy functional
\be\label{energy_func}
H[q_0]=\int_0^\infty d\rho \left[
\frac{(1-q_0^2)^2}{2\rho^2} + \left(\partial_\rho q_0\right)^2 +\rho^2q_0^{2m}
\right].
\ee

For $m\ge 2$ Eq.~(\ref{dif_eq}) is reduced to the equation
\be\label{dif_eq_reduced}
\frac{d^2 q_0}{d\rho^2} + \frac{q_0}{\rho^2}-m\rho^2q_0^{2m-1}=0 \qquad \text{at} \qquad \rho\to \infty,
\ee
which is different from (\ref{dif_eq_asymp_1}) and has the asymptotic solution
\be\label{asymptotic}
q_0=
A\rho^{-\xi}, \quad \text{where} \quad A=\left[\frac{m^2+3}{m(m-1)^2}\right]^{\frac1{2(m-1)}}\quad\text{and}\quad
\xi=\frac2{m-1}\;.
\ee

\underline{For $m=2$}
\be
\begin{aligned}
\label{casem=2}
&\alpha(r)=\arctan(\tilde{\rho}\sqrt{2+\tilde \rho^2}),\\
&\cos\alpha=\frac1{1+\tilde\rho^2}, \quad
\sin\alpha=\frac{\tilde{\rho}\sqrt{2+\tilde \rho^2}}{1+\tilde\rho^2},\quad
\tilde\rho=\sqrt[4]{\tfrac27}\rho.
\end{aligned}
\ee

\underline{For $m=3$}
\be
\alpha(r)=\arctan(\rho),\quad
\cos\alpha=\frac1{\sqrt{1+\rho^2}}, \quad
\sin\alpha=\frac{\rho}{\sqrt{1+\rho^2}},
\ee
see \cite{Manfried}.

For $m\ge 4$ one can connect the two solutions (\ref{small_distances}) and (\ref{asymptotic}) by the following trial function
\be\label{trial_2}
q_0=(1+\kappa_1 \rho^2 +\kappa_2 \rho^4)^{-\xi/4}
\ee
with
\be
\kappa_2=A^{-\frac4{\xi}}=\frac{m(m-1)^2}{m^2+3}
\ee
and $\kappa_1\ge 0$ is a variational parameter.\\[-0.5cm]

\underline{For $m=4$},
$$
\cos\alpha(r) \approx\frac1{\sqrt[6]{1+\kappa_1\rho^2+\kappa_2\rho^4}}
$$
with $\kappa_1=2.98428$, $\kappa_2=\tfrac{36}{19}$.
%

\end{document}